\documentclass[12pt,preprint]{aastex}

\def\HST{{\it HST}}

\def\Teff{T_{\rm eff}}

\shorttitle{M31 RV}
\shortauthors{Bond}

\begin{document}

\title{\emph{Hubble Space Telescope} Imaging of the Outburst Site of
M31~RV. \\
II. No Blue Remnant in Quiescence\altaffilmark{1}}

\author{Howard E. Bond}

\affil{Space Telescope Science Institute, 3700 San Martin Drive, Baltimore, MD
21218; bond@stsci.edu}

\altaffiltext{1}
{Based on observations made with the NASA/ESA {\it Hubble Space Telescope\/},
obtained by the Space Telescope Science Institute, and from the data archive at
STScI. STScI is operated by the Association of Universities for Research in
Astronomy, Inc., under NASA contract NAS5-26555.}

\clearpage

\begin{abstract}

M31 RV is a red transient that erupted in 1988 in the Andromeda bulge, reaching
a luminosity intermediate between novae and supernovae. It was cool throughout
its outburst, unlike a normal classical nova. 

In 2006, Bond \& Siegel examined archival {\it HST\/} optical images of the
M31~RV site, obtained in 1999. We found only old red giants at the site, and no
stars of unusual color. However, Shara et al.\ recently claimed to have detected
(a)~a bright UV source within the error box in {\it HST\/} UV images taken in
1995, (b)~a hot ($\Teff>40,000$~K) optical source in the same 1999 images that
we examined, and (c)~cooling of this source from 1999 to 2008. Shara et al.\
argue that this source's behavior is consistent with a classical-nova outburst
occurring on a low-mass white dwarf.

I have re-examined all of the {\it HST\/} frames, including new ones obtained 
in 2009--2010. I find that: (a)~the bright 1995 UV source reported by Shara
et~al.\ was actually due to cosmic rays striking the same pixel in two
successive exposures; (b)~the claim that an optically bright star in the error
box is very hot is actually due to misinterpretation of red-giant colors in the
STmagnitude system; (c)~there is no evidence for variability of any source
within the error box from 1999 to 2010; and (d)~there are no stars of unusually
blue or red color in the error box. Our 2006 conclusions remain valid: either
M31~RV had faded below \HST\/ detectability by 1999, or its remnant is an
unresolved companion of a red giant in the field, or the remnant {\it is\/} one
of the red giants. 

\end{abstract}


\clearpage

\section{Introduction: Intermediate-Luminosity Red Transients}

Over the past several years, several new classes of outbursting objects have
been recognized, having maximum luminosities intermediate between those of
classical novae and supernovae, and eruptions lasting several weeks to a few
months. These transients typically become very red as their outbursts progress.
I will call them ``intermediate-luminosity red transients'' (ILRTs). They have
alternatively been designated ``luminous red novae'' (e.g., Kasliwal et al.\
2011 and references therein), but this terminology might suggest that the events
belong to a subclass of ordinary novae.

The Galactic variable star V838~Monocerotis is a prototypical ILRT\null. Its
2002 eruption illuminated a spectacular light echo (Bond et al.\ 2003; Bond
2007) from which a geometric distance was obtained (Sparks et al.\ 2008),
implying an absolute magnitude at maximum of $M_V=-9.8$. V838~Mon became very
red during its outburst, eventually enshrouding itself in circumstellar dust.
The mechanism producing its eruption remains controversial, but it has been
argued that it was due to a stellar merger (e.g., Soker \& Tylenda 2006; Corradi
\& Munari 2007, and references therein). 

New support for the stellar-merger hypothesis for at least some ILRTs comes from
the recent discovery (Tylenda et al.\ 2011) that the progenitor of the 2008
Galactic ILRT V1309~Scorpii was a short-period contact binary. Similarities
between the outbursts of V1309~Sco and V838~Mon had been pointed out earlier by
Mason et al.\ (2010).

However, in the past three years, several ILRTs have occurred in external
galaxies, and clearly had an origin in heavily dust-enshrouded massive stars.
Two of the best-studied members of this class are the 2008 optical transient
(OT) in the nearby galaxy NGC~300, designated NGC~300-OT-2008 (Bond et al.\
2009; Berger et al.\ 2009), and SN~2008S in NGC~6946 (Thompson et al.\ 2008;
Smith et al.\ 2008).  Both NGC~300-OT-2008 and SN~2008S were identified with
luminous mid-IR sources in pre-outburst {\it Spitzer Space Telescope\/} images
(Prieto 2008; Prieto et al.\ 2008); however, NGC~300-OT-2008 and SN~2008S were
both extremely faint in the optical band before their eruptions (Bond et al.\
2009; Prieto et al.\ 2008). The outbursts of these dust-obscured objects may
thus be related to the eruptions of luminous blue variables (LBVs), and it is
unclear whether they are connected to lower-luminosity ILRTs such as V838~Mon
and V1309~Sco.  Experiments like the Palomar Transient Factory are now finding
ILRTs in nearby galaxies in large numbers (e.g., Kasliwal et al.\ 2011, and
references therein).  Amateur astronomers are also continuing to contribute
discoveries; the discoverer of NGC~300-OT-2008 recently found a second ILRT in
the same galaxy, designated SN~2010da (Monard 2010), which was also a luminous
mid-IR source before its eruption (Khan et al.\ 2010). These objects are
sometimes called ``supernova impostors,'' having luminosities at maximum several
magnitudes fainter than true supernovae (e.g., Van~Dyk 2005).

\section{M31 RV, the M31 Red Variable}

One of the earliest ILRTs to be recognized had its eruption in 1988. This event
occurred in the nuclear bulge of the Andromeda Galaxy, M31; it was discovered
independently by Rich et~al.\ (1989), Bryan \& Royer (1991), and Tomaney \&
Shafter (1992).  Its maximum luminosity ($M_{\rm bol}\simeq-10$; Rich et al.\
1989) was somewhat brighter than that of the most luminous classical novae, but
the transient remained cool and red throughout its $\sim$3-month eruption. Thus
its behavior was completely different from that of a nova, in which an extremely
hot and blue remnant is quickly revealed as the ejected envelope expands and
becomes optically thin. This transient was, however, rather similar in
luminosity, outburst duration, and color evolution to V838~Mon. This remarkable
object has been called the ``M31 red variable,'' or ``M31~RV\null.'' 
Unfortunately, M31~RV was not observed extensively during its outburst; the
available information is summarized in Bond \& Siegel (2006; hereafter Paper~I)
and Shara et al.\ (2010a; hereafter SZPYK10), and references therein.

M31~RV appears to belong to a subclass of ILRTs that arise from old populations.
Another possible member of this subclass is M85-OT-2006 (Rau et al.\ 2007;
Kulkarni et al.\ 2007), which occurred in an S0 galaxy lacking obvious signs of
massive stars. V1309~Sco, along with another Milky Way ILRT, V4332~Sgr (Martini
et al.\ 1999), also do not appear to be associated with young regions.

In Paper~I we studied M31~RV using archival images obtained with the {\it Hubble
Space Telescope\/} (\HST\/) that fortuitously covered the site of M31~RV\null.
We found no evidence for any stars of unusual color at the outburst site in
images taken in 1999, 11~years after the eruption; within the error box there
are only normal red giants belonging to the old population of the M31 bulge. 
However, SZPYK10 have investigated the same archival material that was available
to us, but reached radically different conclusions.  Specifically, they reported
that there was a UV-bright object at the M31~RV site in archival \HST\/ images
taken in 1995.  Moreover, they found an extremely blue object in the 1999
optical frames that we had examined---the images in which we found only normal
red giants---and they further claimed that the hot source had varied in
additional frames that they obtained in 2008. SZPYK10 therefore argue that
M31~RV may have been an unusual nova occurring on a low-mass white dwarf, rather
than a stellar merger; in an accompanying paper (Shara et al.\ 2010b) they
present theoretical models of classical-nova eruptions that produce slow, red
outbursts, evolving later to high temperatures, similar to the behavior that
they claim for M31~RV\null. If these findings are correct, there may be yet
another evolutionary channel that produces ILRTs and which could explain their
occurrence in old populations.

Still more \HST\/ observations of the M31~RV site have been obtained since the
SZPYK10 study.  In this paper I will critically examine all of the \HST\/
material available up to the present time, with a particular aim of clarifying
the different conclusions reached from the same data by the two different
groups.

\section{\emph{HST} Imaging of the M31 RV Site}

I have searched the \HST\/ data archive\footnote{The \HST\/ data archive is
available at http://archive.stsci.edu/hst} for all imaging observations that
have covered the site of M31~RV\null. My search is complete through 2011
January~1.  The available data are summarized in Table~1. Most of these
observations were made for other purposes and contain the location of M31~RV
only fortuitously, except that the two programs of Shara and myself were
specifically targeted at M31~RV\null.

As described in Paper~I, we located the site of M31~RV in the \HST\/ frames of
1999, by registering them with ground-based CCD images taken during the 1988
outburst by R.~Ciardullo and kindly made available to us. The resulting error
box in the \HST\/ frames has a 1$\sigma$ size of
$\pm$$0\farcs18$$\times$$\pm$$0\farcs27$. SZPYK10 independently determined the location of M31~RV in the \HST\/ frames and verified the Paper~I result to within $0\farcs09$.

\subsection{The 1995 UV Observation: Cosmic Rays Do Strike Twice}

The first \HST\/ images of the M31~RV site were obtained in 1995 (see Table~1).
A pair of 1300~s frames was taken in the WFPC2 F300W filter, a bandpass located
in the optical UV around the atmospheric cutoff. SZPYK10 reported finding a
bright UV source in these frames within the M31~RV error box. Moreover, they
suggested that this source coincides with a relatively bright star seen in
frames taken at longer wavelengths (see below). I will call this star
``Star~S.''  Its J2000 coordinates are 00:43:02.43, +41:12:57.0.

To investigate this claim, which is at variance with what we had reported in
Paper~I, I re-examined the pair of archival images from 1995. The two F300W
exposures were taken in immediate succession (during the same telescope orbit)
without dithering the telescope pointing, so I simply combined the images using
a standard cosmic-ray (CR) rejection algorithm. Because there are very few stars
detected in the F300W frames, I then registered the combined frame with a frame
taken in 2003 at a longer wavelength in which the location of the error box
could be identified unambiguously. The combined F300W image indeed shows an
apparent object within the M31~RV error box, but it has a very unusual
appearance: it looks abnormally sharp compared to images of nearby real stars. 
I measure the FWHM of the object to be only 0.7~pixels, in contrast with a
typical 1.4--1.5~pixels for real stars in the frame. The FWHM of stellar images
is set by diffraction in the telescope optics, and it is physically impossible
for a real point source to have a FWHM of 0.7~pixels.

Figure~1 illustrates the unusual visual appearance of the SZPYK10 object in the
two individual F300W images. I created this mosaic by extracting $9\times9$
pixel ($0\farcs9$$\times$$0\farcs9$) postage stamps around the object (the two left-hand frames in the mosaic) and four real stars from the same two images. (The real stars were selected to be free of CR hits within the $9\times9$
boxes.) It is immediately clear that the SZPYK10 object lacks the broad PSF
wings associated with real stars in these images. The same image stretch was
used throughout Figure~1, and the real stars cover a range of magnitudes that
bracket the ``magnitude'' of the M31~RV candidate. Thus the obvious difference
in appearance between the candidate and the real stars is not an artifact of the
image presentation.

To quantify this disparity, I measured two parameters in the two individual
F300W frames for the SZPYK10 object and several real stars: the signal in data
numbers (DN) for the brightest pixel in the image, and the total
background-subtracted flux (also in DN) within a 2-pixel-radius aperture
centered on the object. Figure~2 plots the results. Note that I made these
measurements on raw frames, which have a bias level of about 355~DN\null. Only
stars without CR hits in their vicinity were measured; there are actually very
few uncontaminated real stars in these frames, and essentially all of them are
plotted in Figure~2.  There is, as expected, a linear relation between the DN
value of the brightest pixel and the total stellar flux (with some scatter due
to different centerings of the stars within the central pixel as well as
Poissonian noise); a least-squares linear fit to these values is plotted in
Figure~2. However, the SZPYK10 object stands out strikingly from this relation
in both individual images: it has an unusually bright central pixel relative to
the total stellar flux in the image. In other words, {\it most of the flux of
its image is contributed by the central pixel,} and this is true for {\it
both\/} individual images. The images lack the appreciable extended PSF wings of
real stars.  When a CR deposits energy in a single WFPC2 CCD pixel, about 25\%
the charge diffuses to the immediately adjacent pixels (see Riess, Biretta, \&
Casertano 1999). The dashed line in Figure~2 plots the predicted behavior of
single pixels impacted by CRs, with 75\% of the charge remaining in the central
pixel.  It provides a good fit to the two measurements of the SZPYK10 object.

In summary, the candidate object has a visual appearance very different from
that of real stars, and quantitatively its image parameters not only depart
significantly from those of real stars, but have parameters that are predictable
from the charge-diffusion properties of the detector. The conclusion seems
inescapable that the apparent object in these frames is in fact a case of CRs
striking the same pixel in both images (in the second frame, the CR actually
spread its charge over two pixels), and thus not being removed from a combined
image made with a standard CR-rejection algorithm.  

How probable is it that CRs would hit the same pixel in two successive
exposures? The brightest pixel in the first frame showing the SZPYK10 object has
a value of 514~DN, and in the second frame the two brightest pixels have values
of 502 and 398~DN\null. The fraction of pixels having CR hits of 398~DN or
brighter in the two entire images is measured to be 1.1\%.  The total number of
pixels within the 3$\sigma$ M31~RV error box is 44, so the probability of a CR
hit somewhere in the error box in a single image is $\sim$48\%.  The probability
of this same pixel then being hit again in the second image is 1.1\%, a small
value but not vanishingly so.

The site of M31~RV was imaged again in the same WFPC2 F300W filter in 2008 (see
Table~1). This time, as reported by SZPYK10, and verified independently by
myself, no UV source was detected anywhere within the M31~RV error box. There
were four individual dithered frames taken for this observation, in contrast to
only two in 1995, resulting in a negligible number of CR hits surviving into the
combined frame. Also, the long (10,800~s) WFPC2 exposure of the site taken in
1995 with the F170W near-UV filter shows nothing at the M31~RV site, as reported
by SZPYK10 and confirmed by myself. 

This is not the first time that CRs striking the same pixel in WFPC2 images have
resulted in an astrophysical misinterpretation: see Sahu, Anderson, \& King
(2002).

\subsection{Star S: Not so Hot}

As noted above, SZPYK10 reported that the purported UV object in the 1995 frames
coincides with a bright star, Star~S, seen in images taken at longer
wavelengths. Fig.~3 in SZPYK10 gives a finding chart for this star, and it is
also visible (but not marked) in Fig.~1 of Paper~I\null. Although I have just
presented evidence that the 1995 object is a spurious case of CRs striking
twice, I carried out a careful registration of the 1995 frames with an ACS image
obtained in an F435W filter in 2003, using stars detected in both images. The
result shows that the spurious UV object is offset from the real optical Star~S
by $0\farcs17$, approximately 11~times the 1$\sigma$ error of the registration.
Thus, even if the UV source had been real, it did not coincide with the star
seen at optical wavelengths.

However, SZPYK10 have further reported that the optical star is blue and
extremely hot, and that it is therefore a strong candidate for the remnant of
M31~RV as well as providing support for the classical-nova hypothesis. This is
again in striking contradiction to the findings we reported in Paper~I\null.  In
Figure~3 of the present paper, I have reproduced our color-magnitude diagram
(CMD) from Paper~I, which plots stars that lie within the 3$\sigma$ positional
error box for M31~RV\null. This CMD was derived by M.~Siegel from PSF photometry
of an excellent set of dithered WFPC2/PC frames obtained in 1999, as described
in detail in Paper~I\null. The magnitudes in the WFPC2 F555W and F814W filters
were transformed to the traditional ground-based Johnson-Kron-Cousins $V$ and
$I$ system. Star~S is the brightest object in the error box, and is now labelled
in Figure~3. We had measured its magnitudes and color as $V=22.95$, $I=21.33$, and $V\!-\!I=1.62$, with errors of about $\pm$0.01--0.02~mag. As we already
remarked in Paper~I, this CMD shows only a population of cool red giants
belonging to the M31 bulge. 

To verify that there are only old red giants in the error box, I have also added
three isochrones to Figure~3, taken from the BaSTI database\footnote{Available
at http://albione.oa-teramo.inaf.it/} (Pietrinferni et al.\ 2004, 2006). I chose
parameters typically found in recent studies of the M31 bulge population (e.g.,
Saglia et al.\ 2010 and references therein): an age of 10~Gyr and metallicities
of $\rm [M/H] = +0.25$, $-0.25$, and $-0.66$, with scaled solar element ratios.
I then adjusted the isochrones for a reddening of $E(B\!-\!V)=0.062$ (Schlegel,
Finkbeiner, \& Davis 1998), corresponding to $E(V\!-\!I)=0.08$, and for an M31
distance modulus of $(m-M)_0=24.4$ (van~den~Bergh 2000). The three adjusted
isochrones nicely bracket the positions of all of these normal red giants in the
CMD\null.

Star~S, with a Johnson-Kron-Cousins color index of $(V\!-\!I)_0 = 1.54$, has an
implied effective temperature near $\Teff=4000$~K (e.g., Bessell 1979). How then
did SZPYK10 find $(V\!-\!I)_0\simeq-0.35$ for the same star from the same 1999 data, and then derive a temperature of $\Teff>40,000$~K\null? The discrepancy is partly due to the considerably higher reddening of $E(V\!-\!I)\simeq0.4$ adopted by SZPYK10, compared to the $E(V\!-\!I)\simeq0.08$ used in most studies of the M31 bulge (and verified by the good isochrone fit in Figure~3). However, most of the difference appears to arise from a misunderstanding of the ``STmagnitude'' system that is sometimes used in analyses of \HST\/ data. It is clear, e.g.\ from the axis labels in their figures, that SZPYK10 used STmagnitudes.

STmagnitudes are easily obtained from \HST\/ images through calibration
information included in the image headers, but it is essential to understand
their meaning. They are defined (e.g., Sirianni et al.\ 2005) in terms of
stellar flux density, $F_\lambda$, through the formula ${\rm STmag} = -2.5\log
F_\lambda + \rm const.$, where the constant is set such that the $V_{\rm STmag}$
of Vega is zero. By contrast, a hypothetical object with a flat $F_\lambda$
energy distribution would have colors of zero in the STmagnitude system. The
Johnson-Kron-Cousins $V-I$ color index of Vega in the STmagnitude system is
$(V-I)_{\rm STmag}=-1.27$, not zero. If STmagnitude-based $V-I$ colors are
interpreted using calibrations, such as that of Bessell, based on a
Vega-magnitude system, stellar temperatures will be grossly overestimated. This
appears to explain why Star~S was misinterpreted as a very blue and hot star,
when in fact it is an ordinary red giant.

\subsection{Is Star S Variable?}

Lastly, SZPYK10 report that the candidate M31~RV remnant, Star~S in my
terminology, has varied at longer wavelengths, becoming redder between 1999 and
2008. I will report detailed PSF photometry of all objects in the error box in a
separate paper, but I have made an exploratory investigation of the star's
variability using simple quick-look aperture photometry. I measured the
brightness of Star~S within a 2-pixel-radius aperture in all of the available
frames taken in F435W, F475W, F555W, and F814W\null.   Unfortunately, a wide
variety of cameras has been used in these observations, making direct
comparisons difficult.  Nevertheless, the results are presented in Table~2.
Star~S was, at best, only marginally detected in all filters shortward of F435W,
so those data are omitted. The magnitudes listed in Table~2 are differential
with respect to a nearby, well-isolated comparison star (J2000 coordinates:
00:43:02.56, +41:12:58.3).  

Table~2 shows little evidence for variability of Star~S exceeding about
$\pm$0.1~mag, which is the approximate precision of these quick-look
measurements, over the interval from 1999 to 2010. Note also the similarity of
the differential magnitudes across all of the filters; this shows that the
comparison star has very similar colors to Star~S, and is another normal red
giant of the M31 bulge.

\section{Are There Other Candidate Remnants?}

\subsection{Preliminary Search for Variable Sources}

I have made a preliminary search for other possible candidate remnants of M31~RV
by registering frames taken in the same bandpasses, and then visually blinking
them as well as calculating difference images. This was done using the five
F814W frames (1999 to 2010), and separately using the three F555W frames (1999
to 2009), as listed in Table~1. Although there are many variable stars in these
frames, none of them lie within the M31~RV error box. For well-exposed stars,
variability in excess of $\sim$0.2~mag would have been seen. A separate search
for variable objects at fainter levels will be reported later, but there are no
objects among the numerous fainter sources that appear to have faded
dramatically or disappeared.

\subsection{Near-Infrared Images}

The M31~RV site has been imaged once by \HST\/ in the near-IR, using the IR
channel of WFC3 and the broad-band F110W and F160W filters. I have compared
these frames visually with the optical (F555W and F814W) images, and found no
unusually red objects. Thus, if the remnant of M31~RV is now embedded in dust,
the obscuration is such that the source is not detectable out to the $H$ band.

\section{Summary}

I have investigated all available \HST\/ images of the site of M31~RV, an ILRT
event that occurred in M31 in 1988. These frames were taken between 1995 and
2010. I have been unable to verify the reports by SZPYK10 of (a)~a UV-bright
object at the outburst site in 1995 (the apparent source is due to an
unfortunate case of cosmic rays striking the same pixel in two different
images); (b)~a hot optical source in 1999 and 2008 (due to a misinterpretation
of the \HST\/ STmagnitude system); and (c)~variability of this candidate
source.  

Thus the conclusions we reached in Paper~I all remain valid. I quote them
verbatim: ``All of the stars at the outburst site have the magnitudes and colors
of ordinary red giants in the M31 bulge. The absence of any conspicuous remnant
star has three possible explanations: (a)~the object had faded below \HST\/
detectability in the 11~years since outburst, either intrinsically or because of
heavy dust obscuration; (b)~the remnant is an unseen companion of (or its image
is blended with) one of the red giants in the field; or (c)~the remnant {\it
is\/} one of the red giants.'' In addition, none of the red giants within the
error box have faded significantly over the 1995-2010 time interval covered by
the available \HST\/ images. The nature of M31~RV remains as puzzling as ever.

\acknowledgments

I thank Jay Anderson, Stefano Casertano, and Ron Gilliland for informative
discussions of cosmic-ray hits in WFPC2 frames, Jos\'e Prieto for discussions of
ILRTs, and Mike Siegel for his contributions to Paper~I\null. This work has made
use of the BaSTI web tools that provide stellar isochrones. Support for this
work was provided by NASA through grant number GO-11716 from the Space Telescope
Science Institute, which is operated by AURA, Inc., under NASA contract NAS
5-26555.

{\it Facilities:} \facility{\HST\/ (WFPC2, ACS, WFC3)}

\clearpage

\begin{figure}
\begin{center}
\includegraphics[width=4in]{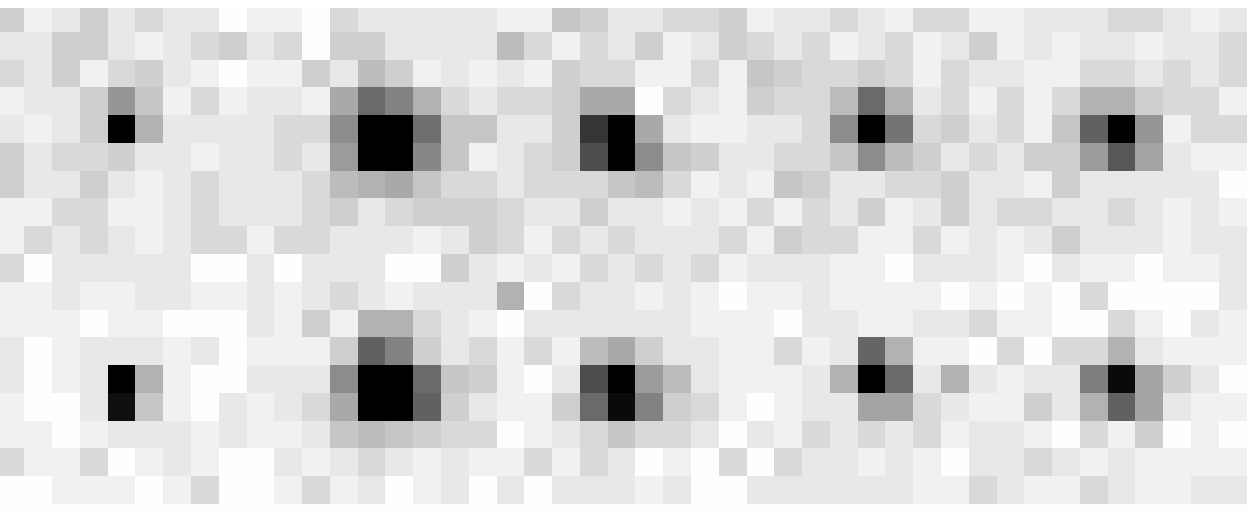}
\end{center} 
\figcaption{A mosaic of $9\times9$ pixel postage stamps extracted from two WFPC2
frames obtained in 1995 in the F300W ultraviolet filter. The two postage stamps
on the left show the SZPYK10 candidate object (``Star~S'') lying within the
M31~RV error box, and the rest of the images are of four real stars from the
same frames selected to be free of cosmic-ray hits. The same image stretch is
used throughout, and the four stars have magnitudes bracketing the ``magnitude''
of the candidate object. The SZPYK10 object is dominated by one or two bright
pixels, and lacks the fainter PSF wings that surround real stars in \HST\/
images.}
\end{figure}

\begin{figure}
\begin{center}
\includegraphics[width=6in]{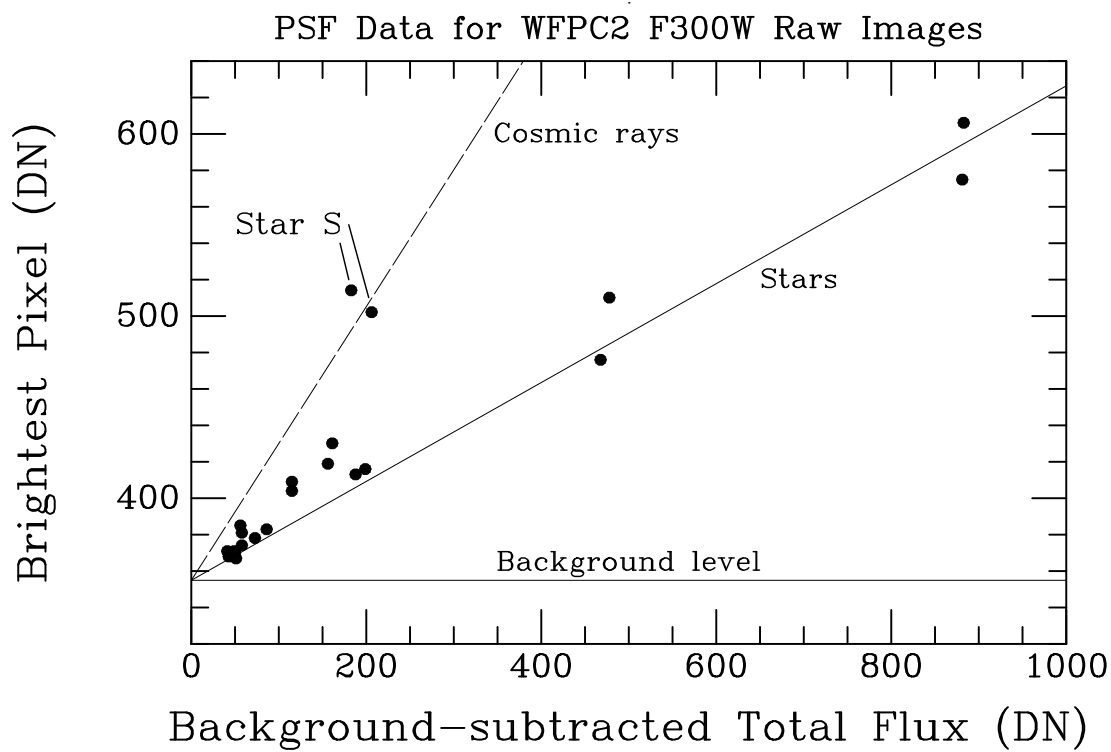}
\end{center} 
\figcaption{Plot of the brightness of the brightest pixel in a sample of stellar
images in two F300W frames vs.\ the total background-subtracted flux within a
2-pixel aperture. The {\it solid line\/} is a least-squares fit to the real
stars, and the {\it dashed line\/} is the predicted relation for cosmic-ray hits
where all of the energy is deposited in a single pixel, with subsequent charge
diffusion to neighboring pixels. Star~S stands out from all of the real stars in
having most of its flux contributed by the one or two central pixels, and this
is true in both individual frames. This indicates that the apparent stellar
object is actually a case of cosmic rays striking the same pixel in both images.
The horizontal line at the bottom indicates the bias level of 355~DN in these
raw frames.}
\end{figure}

\begin{figure}
\begin{center}
\includegraphics[width=6in]{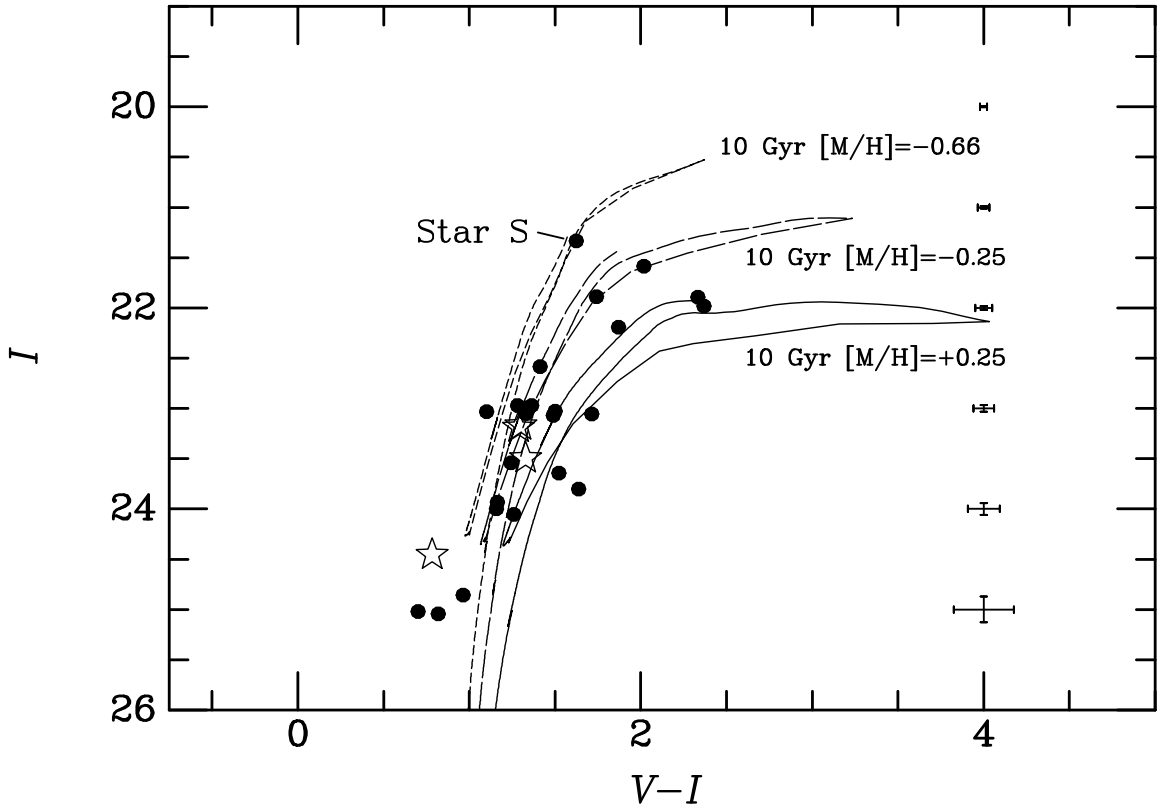}
\end{center} 
\figcaption{Color-magnitude diagram for stars lying within a radius of 6~WFPC2
pixels ({\it stars\/}) and within 6 to 18~pixels ({\it filled circles\/}) of the
M31~RV site, taken from Paper~I\null.  These radii correspond to locations
within 1$\sigma$ and 3$\sigma$ of the outburst site (see text); the
corresponding angular radii are $0\farcs27$ and $0\farcs82$. The bright
candidate Star~S is labelled. Also plotted are BaSTI isochrones for an age of
10~Gyr and metallicities of $\rm[M/H]=-0.66$, $-0.25$, and $+0.25$, adjusted to
the reddening and distance of M31.  All of the stars within the error box are
normal M31 old red giants, including Star~S\null. Error bars at the right-hand
side show the mean photometric errors as functions of $I$ magnitude.}
\end{figure}

\clearpage


\begin{deluxetable}{lllcl}
\tablewidth{0 pt}
\tablecaption{\HST\/ Imaging of the M31 RV Outburst Site}
\tablehead{
\colhead{Date (UT)} &
\colhead{Camera} &
\colhead{Filter} &
\colhead{Total Exp.} &
\colhead{Program ID/PI} \\
\colhead{} &
\colhead{} &
\colhead{} &
\colhead{Time (s)} &
\colhead{} 
}
\startdata
1995 December 5    & WFPC2/WF & F170W & \llap{1}0800 & 6255/King  \\
                   &          & F300W & 2600         &               \\
\noalign{\smallskip}
1999 July 23-24    & WFPC2/PC & F555W & 7200         & 8018/Green \\	
                   &          & F814W & \llap{1}0400 &               \\	
\noalign{\smallskip}
2003 December 25   & ACS/WFC  & F435W & 2200         & \llap{1}0006/Garcia \\
2004 October 2     &          & F435W & 2200         & \llap{1}0006/Garcia \\
\noalign{\smallskip}
2008 July 26       & WFPC2/PC & F300W & 2400         & \llap{1}1546/Shara \\
                   &          & F439W & 1600         &                    \\
                   &          & F555W & 800          &                    \\
                   &          & F814W & 2400         &                    \\
\noalign{\smallskip}
2008 July 27       & ACS/SBC  & F140LP & 2552        & \llap{1}1546/Shara \\
\noalign{\smallskip}
2009 December 7    & WFC3/UVIS & F555W & 1707        & \llap{1}1716/Bond \\
                   &           & F814W &  987        &            \\
\noalign{\smallskip}
2010 July 23       & ACS/WFC   & F475W & 1720        & \llap{1}2058/Dalcanton \\
                   &           & F814W & 1520        &  \\
\noalign{\smallskip}
2010 July 25       & ACS/WFC   & F475W & 1720        & \llap{1}2058/Dalcanton \\
                   &           & F814W & 1520        &  \\
\noalign{\smallskip}
2010 December 24   & WFC3/UVIS & F275W & 1010        & \llap{1}2058/Dalcanton \\
                   &           & F336W & 1350        &  \\
\noalign{\smallskip}
2010 December 25-26 & WFC3/IR  & F110W &  799        & \llap{1}2058/Dalcanton \\
                    &          & F160W & 1697        &  \\
\noalign{\smallskip}
2010 December 26   & WFC3/UVIS & F275W & 1010        & \llap{1}2058/Dalcanton \\
                   &           & F336W & 1350        &  \\
\enddata
\tablecomments{Camera abbreviations are: WFPC2 = Wide Field Planetary Camera~2
(WF = wide-field chip, PC = planetary chip); ACS = Advanced Camera for Surveys
(WFC = wide-field optical channel; SBC = solar-blind UV channel); WFC3 = Wide
Field Camera~3 (UVIS = UV/optical channel; IR = near-infrared channel). }
\end{deluxetable}

\begin{deluxetable}{llcccc}
\tablewidth{0 pt}
\tablecaption{Relative Photometry (In Magnitudes) of Star S}
\tablehead{
\colhead{Date} &
\colhead{Camera} &
\colhead{$\Delta(F435W)$} &
\colhead{$\Delta(F475W)$} &
\colhead{$\Delta(F555W)$} &
\colhead{$\Delta(F814W)$}}
\startdata
1999 July 23     & WFPC2 &  $\dots$ & $\dots$ & 0.59	& 0.46	  \\
2003 December 25 & ACS   &  0.62    & $\dots$ & $\dots$ & $\dots$ \\
2004 October 2   & ACS   &  0.56    & $\dots$ & $\dots$ & $\dots$ \\
2008 July 26     & WFPC2 &  $\dots$ & $\dots$ & 0.50	& 0.38	  \\
2009 December 7  & WFC3  &  $\dots$ & $\dots$ & 0.57	& 0.46	  \\
2010 July 23     & ACS   &  $\dots$ & 0.69    & $\dots$ & 0.57	  \\
2010 July 25     & ACS   &  $\dots$ & 0.52    & $\dots$ & 0.52	  \\
\enddata
\end{deluxetable}

\end{document}